\documentstyle[12pt,epsf]{article}
\begin{document}
\def\doublespaced{\baselineskip=\normalbaselineskip\multiply
    \baselineskip by 150\divide\baselineskip by 100}
\doublespaced
\pagenumbering{arabic}
%
\begin{titlepage}
\begin{flushright}
{ June, 1996 }
\end{flushright}
\vspace{2cm}
\begin{center} 
\large
{\bf Effects of QCD Resummation on Distributions of\\
  Top--Antitop Quark Pairs Produced at the Tevatron }
\end{center}
\vspace{0.4cm}
\begin{center}
{\bf  S. Mrenna~$^{(a),}$\footnote{mrenna@hep.anl.gov}~~~ and
~~~C.--P. Yuan~$^{(b),}$\footnote{yuan@msupa.pa.msu.edu}}
\end{center}
\vspace{0.2cm}
\begin{center}
{
(a) High Energy Physics Division, Argonne National Laboratory \\
Argonne, IL 60439, U.S.A. \\
(b) Department of Physics and Astronomy,
Michigan State University \\
East Lansing, MI 48824, U.S.A. \\
}
\end{center} 
\vspace{0.4cm}
\raggedbottom
\setcounter{page}{1}
\relax

\begin{abstract}
\noindent
We study the kinematic distributions of top--antitop quark ($t\bar t$) pairs 
produced at the Tevatron, 
including the effects of initial state and final state 
multiple soft gluon emission, using the Collins--Soper--Sterman resummation 
formalism. The resummed results are compared with those predicted by
the showering event generator {\tt PYTHIA}
for various distributions involving the $t\bar t$ pair and the
individual $t$ or $\bar t$.  The comparison between the experimental and
predicted distributions will be a strong test of our understanding and
application of perturbative QCD.
Our results indicate that the showering
event generators do not produce enough radiation.
We reweight the {\tt PYTHIA}
distributions to agree with our resummed calculation, then use the
reweighted events to better estimate the true hadronic activity in
$t\bar t$ production at hadron colliders.
\end{abstract}
\end{titlepage}
\newpage
\section{Introduction}
\indent

Because the top quark mass $m_t$ is comparable in magnitude
to the vacuum expectation value $v$=246 GeV \cite{top}, 
studying the interactions of the top quark
may provide information on the mechanism of electroweak 
symmetry breaking \cite{ehab} or the generation of fermion masses \cite{sekh}. 
To observe any new physics effect in the top quark
system, one has to know first the Standard Model prediction for
the production rate and the kinematics of the top quarks produced 
at colliders. 
We concentrate on top quark pairs produced in
hadron collisions. 
The next-to-leading-order (NLO) prediction for the production rate of
$t \bar t$ pairs has been known for several years \cite{nlotop}.
Since then, several studies \cite{smith,berger,mangano}
have extended this result to include
the effect of soft gluon radiation on the production rate
of $t \bar t$ pairs at hadron colliders.
The NLO prediction of the $t \bar t$ production rate varies from 
the leading order (LO) prediction by 15--35\% at the Tevatron 
for $m_t$ = 175\,GeV,
depending on the choice of scale for the hard scattering process, and
is less sensitive ($\simeq 10\%$) to the scale choice.
The resummation of multiple soft gluon emission 
may increase the production rate by another 10\%, depending on the prescription 
for performing the resummation \cite{berger,mangano}.
Besides testing the production rate, it is also important to study the
kinematics of the top quark to probe possible new physics 
associated with its production or decay.

It is well established that the transverse momentum $Q_T$ distribution 
of the $t \bar t$ pair cannot be described by the NLO perturbative 
calculation for small $Q_T$.  The same is true for the NLO prediction of
the transverse momentum of electroweak gauge 
bosons \cite{altarelli}.  This implies that
the transverse momentum $p_T^t$ of the top quark cannot
be accurately predicted by the NLO calculation,
especially for $t \bar t$ pairs with small $Q_T$,
where the data dominate.
The effects of the initial state and the final state
multiple soft gluon emission
must be resummed to predict the kinematic distributions of the top quarks 
produced in $t \bar t$ events at hadron colliders.
This work expands upon an earlier study of the kinematics of heavy quark
pairs \cite{berger_meng} using the
Collins--Soper--Sterman formalism to perform the
resummation \cite{collins}. 
We closely follow the notation used 
in Ref.~\cite{sterman}.

Our present understanding of the $t \bar t$ pair kinematics is based on
showering event generators, such as {\tt HERWIG, ISAJET} and {\tt
PYTHIA}~\cite{generators}.  A further goal is to quantify the successes and
limitations of such generators and to make progress towards a more 
complete description of the $t \bar t$ pair and individual $t$
and $\bar t$ kinematics~\cite{frixione}.
This will have important implications for the precision measurement
of the top quark mass.

Following this introduction, we have organized this study into five
additional sections.
Sec.~2 contains a review of the 
Collins--Soper--Sterman (CSS) resummation formalism. In Sec.~3, we
present our numerical results for the $q \bar q \rightarrow t \bar t$ 
and $gg \rightarrow t \bar t$ subprocesses using this formalism.
We compare our results with the showering event generator 
{\tt PYTHIA} in Sec.~4.  Based on the results of Sec.~4, an improved
estimate of the hadronic activity in $t \bar t$ events is presented
in Sec.~5.  Finally, Sec.~6 contains our conclusions.


\section{The CSS Resummation Formalism}
\indent

Soft gluon resummation has been applied successfully to predict
the rate and kinematics of electroweak gauge boson production at hadron
colliders \cite{altarelli,sterman,arnold}.
Although the applicability of the 
CSS formalism to $t \bar t$ pair production (which is a colored
final state) has not been proven in the literature, 
the large top quark mass relative to $\Lambda_{QCD}$ should suppress
contributions from color configurations not included
in the CSS resummation formalism.  This should be more correct
when the $t$ (and $\bar t$) are produced in the central rapidity
region.
All the leading and
sub--leading logarithmic singularities associated with the initial
state radiation in the NLO expression for $q {\bar q}\rightarrow t{\bar
t}$ production are universal to those found for electroweak gauge boson
production \cite{berger_meng}.  
For $gg\rightarrow t{\bar t}$ production, they are the same
as those for Higgs boson production \cite{cphiggs,kauffman_higgs}.
Obviously, there are also singularities associated with the final state
radiation in $t\bar t$ production which are absent in either electroweak
gauge boson or Higgs production.

Our starting point for applying the CSS formalism to $t\bar t$
production at hadron colliders is the resummed expression
for the differential cross section:
  \begin{eqnarray} & &
  \left( { d \sigma(h_1 h_2 \rightarrow t\overline{t}+X)
  \over dQ^2 \, dy \, dQ^2_T \, d\phi_{t \bar t} \, d\cos{\theta} \, d\phi}
  \right)_{res} = 
  { \pi^2 \over 36 S Q^2} \,
  \nonumber \\  & & ~~
  \times \bigg\{ { 1\over (2 \pi)^2}
  \int_{}^{} d^2 b \, e^{i {\vec Q_T} \cdot {\vec b}} \, 
  \sum_{j,k}{\widetilde{W}_{jk} 
   (b_*,Q,x_1,x_2, \theta, \phi,C_1,C_2,C_3)} \,
  F^{NP}_{jk} (b,Q,x_1,x_2) 
  \nonumber \\  & & ~~~~
  + ~  Y(Q_T,Q,x_1,x_2, \theta, \phi,C_4) \bigg\}.
  \label{ResFor}
  \end{eqnarray}
In this expression, the production rate is described in terms of the 
mass $Q$, rapidity $y$,
transverse momentum $Q_T$, and azimuthal angle $\phi_{t \bar t}$ of the
$t \bar t$ pair in the laboratory frame, and the polar angle $\theta$
and azimuthal angle $\phi$ in a special center--of--mass frame for
the $t \bar t$ pair, the Collins--Soper frame
\cite{CSFrame}.
The center-of-mass energy $\sqrt{S}$ of hadrons $h_1$ and $h_2$ fixes
the parton momentum fractions
$x_1 = {Q \over \sqrt{S}}e^{y}, x_2 = {Q \over \sqrt{S}}e^{-y}$.
The renormalization group invariant $\widetilde{W}_{jk}$ is given by
  \begin{eqnarray} &&
  \widetilde{W}_{jk} (b,Q,x_1,x_2, \theta, \phi,C_1,C_2,C_3)  = 
  \exp \left\{ -S(b,Q,C_1,C_2) \right\} \nonumber \\ && ~
  \times  \left[  
  \left( C_{jl} \otimes f_{l/h_1} \right) (x_1) ~
  \left( C_{km} \otimes f_{m/h_2} \right) (x_2) +
  \left( C_{kl} \otimes f_{l/h_1} \right) (x_1) ~
  \left( C_{jm} \otimes f_{m/h_2} \right) (x_2) \right]  
  \nonumber \\ && ~~~~~
  \times \left[{\alpha_s(C_2 Q) \over \pi}\right]^2 
   \left[1+{\alpha_s(C_2 Q) \over \pi}2\beta_1\ln\left({(C_2Q)^2 \over
m_t^2}\right)\right]
   {\cal H}_{jk}(Q,\cos \theta,m_t),
  \label{WTwi}
  \end{eqnarray}
where $\alpha_s$ is the strong coupling constant,
$\beta_1 = {1 \over 12} (33-2n_f)$ ($n_f$ is the number of light quark
flavors) and $\otimes$ denotes the convolution integral
  \begin{eqnarray} & &
  \left( C_{jl} \otimes f_{l/h_1} \right) (x_1) = 
  \int_{x_1}^{1} {d \xi_1 \over \xi_1} \, 
  C_{jl}( {x_1 \over \xi_1}, b, \mu={C_3 \over b}, C_1,C_2)
  f_{l/h_1}(\xi_1, \mu={C_3 \over b}).
  \label{Convol}
  \end{eqnarray}
Because there are two separate hard processes in the LO calculation,
there are two functions $\widetilde{W}_{q \bar q}$ and $\widetilde{W}_{gg}$. 
The dummy indices $l$ and $m$ 
are meant to sum over quarks and anti-quarks or gluons, and
summation on double indices is implied.
The angular function ${\cal H}_{jk}(Q,\cos \theta,m_t)$ in 
Eq.~(\ref{WTwi}) for $j=q, k=\bar q$ is
  \begin{eqnarray} & &
       \left [ 2- \beta^2 + \beta^2 \cos^2\theta \right],
   \nonumber 
  \end{eqnarray}
and for $j=k=g$ is
  \begin{eqnarray} & &
    { 3 \, ( 7 + 9 \beta^2 \cos^2\theta)  \over 
     32 \,(1-\beta^2 \cos^2\theta )^2 }
     \left[ 1+2 \beta^2 - 2 \beta^4 - 2 \beta^2(1-\beta^2)\cos^2\theta 
      - \beta^4 \cos^4\theta \right], 
  \nonumber
  \end{eqnarray}
where $\beta=\sqrt{1 - 4 m_t^2/Q^2 }$.
We generically refer to $\widetilde{W}_{jk}$ as the CSS piece.
The Sudakov form factor $S(b,Q,C_1,C_2)$ is defined as
  \begin{eqnarray} & &
  S(b,Q,C_1,C_2) =
  \int_{C_1^2/b^2}^{C_2^2Q^2}
  {d {\bar \mu}^2\over {\bar \mu}^2}
       \left[ \ln\left({C_2^2Q^2\over {\bar \mu}^2}\right)
        A\big(\alpha_s({\bar \mu})\big) +
        B\big(\alpha_s({\bar \mu})\big)
       \right]. 
  \label{SudExp}
  \end{eqnarray}
The functions $A$, $B$ and the  Wilson coefficients $C_{jl}$  
were given in Ref.~\cite{sterman} for  $q \bar q \rightarrow t \bar t$
(see Eqs.\,(3.19) to (3.26) for 
$A^{(1)}$, $B^{(1)}$, $A^{(2)}$, $C^{(0)}_{jk}$, $C^{(1)}_{jk}$,
and $C^{(1)}_{jg}$), 
and in Ref.~\cite{cphiggs} for  $gg \rightarrow t \bar t$
(see Eqs.~(3.8) to (3.9) for 
$A^{(1)}$, $B^{(1)}$, $C^{(0)}_{gg}$, $C^{(1)}_{gg}$,
and $C^{(1)}_{gq}$).\footnote{The superscripts (0), (1), and (2)
represent the order in $\alpha_s$.  $A,B$ and $C_{jk}$ are all
calculated in the $\overline{\rm MS}$ (modified minimal subtraction) scheme.}
In those results, the constants 
$C_1$, $C_2$ and $C_3 \equiv \mu b$ were introduced
when solving the renormalization group equation for the CSS piece
$\widetilde{W}_{jk}$.
The canonical choice of these renormalization constants is
$C_1 = C_3 = 2 e^{-\gamma_E} \equiv b_0 $ and $C_2 = 1$~
\cite{sterman,cphiggs}.
($\gamma_E$ is the Euler constant.)
To test the dependence of our numerical results on the particular 
choice of the renormalization constants, we consider the set of constants 
such that
$C_1=C_2 b_0$ and $C_3=b_0$. 
This choice eliminates large constant factors in the
expressions for the $A, B$ and $C_{jk}$ functions.
Because the $t \bar t$ final
state is colored, there is an additional contribution to
the $B$ function inside the Sudakov factor due to final state gluon 
radiation. 
The mass of the top quark regulates a potential collinear singularity
so that the final state contributes
only $\ln^{[1]}({Q^2 \over Q_T^2})$ terms due to soft gluon
emission.  Since there are
no $\ln^{[2]}({Q^2 \over Q_T^2})$ contributions at NLO, there is no
$A^{(1)}_{final}$ function.
The additional contribution to the $B$ function was 
given originally in Ref.~\cite{berger_meng},
  \begin{eqnarray}
   B^{(1)}_{final}= C_F \left[
  1+{1+\beta^2 \over \beta} \ln \left( {1-\beta \over 1+\beta} \right)
  \right]. 
  \end{eqnarray}
Near threshold, when $\beta \rightarrow 0$, $B^{(1)}_{final}= - C_F$ with
$C_F=4/3$ in QCD.

As shown in Eq.~(\ref{SudExp}), the upper limit of the integral
for calculating the Sudakov factor is ${\bar \mu}=C_2 Q$, which sets
the scale of the hard scattering process when evaluating
the renormalization group invariant
quantity $\widetilde{W}_{jk}$, as defined in Eq.~(\ref{WTwi}).
The lower limit ${\bar \mu}\equiv C_1/b = b_0/b$ determines the onset of
non--perturbative physics.

The $Y$--term in Eq.~(\ref{ResFor}) is defined as
  \begin{eqnarray}
  Y(Q_T,Q,x_1,x_2,\theta,\phi,C_4) = 
  \int_{x_1}^{1} {d \xi_1 \over \xi_1}
  \int_{x_2}^{1} {d \xi_2 \over \xi_2}
  \sum_{N=1}^{\infty} \left[{\alpha_s(C_4 Q) \over \pi} \right]^{(2+N)} 
  \nonumber \\
  \times f_{l/h_1}(\xi_1;C_4 Q) \, R_{lm}^{(N)} (Q_T,Q,{ x_1 \over \xi_1},
  { x_2 \over \xi_2},\theta,\phi)
  \, f_{m/h_2}(\xi_2;C_4 Q) ,
  \label{RegPc}
  \end{eqnarray}
where the functions $R_{lm}^{(N)}$ 
only contain contributions less singular than 
${1 \over Q_T^{2}} \times(1 \, {\rm or} \, \ln({Q^2 \over Q_T^2}))$ 
as $Q_T \rightarrow 0$.
We denote those singular contributions as the singular--piece
in contrast to the regular $Y$--piece.  
The scale of the $Y$--piece is specified by the choice of $C_4$.
To {\it optimize} the 
perturbative expansion, that is, to {\it minimize} the contribution
of logarithmic terms $\ln(C_4)$ from
higher order corrections, we choose 
$C_4=1$ in calculating the $Y$--piece.
More specifically, to obtain the regular 
$Y$--piece, we subtract the singular-piece  
for $ q \bar q \rightarrow t \bar t g$,
$ g q/{\bar q} \rightarrow t \bar t q/{\bar q}$, and
$ gg  \rightarrow t \bar t g$
(which can be obtained by expanding Eq.~1 to order $\alpha_s^3$
with $C_1 = C_2 b_0$
and retaining those terms proportional to $Q_T^{-2}$)
from the squared amplitude  for the tree level processes
$ q \bar q \rightarrow t \bar t g$,
$ g q/{\bar q} \rightarrow t \bar t q/{\bar q}$, and
$ gg  \rightarrow t \bar t g$.

In Eq.~(\ref{ResFor}), the impact parameter $b$ is to be integrated 
from 0 to $\infty$. 
However, for $b \ge b_{max}$, which corresponds to an energy scale 
less than $1/b_{max}$, the 
QCD coupling $\alpha_s$ becomes so large that a perturbative 
calculation is no longer reliable.\footnote{
We use $b_{max}=0.5\,{\rm GeV}^{-1}$ in our calculation. 
}
The non-perturbative function 
$F^{NP}$ is needed in the formalism with the general
structure
  \begin{eqnarray}
  F^{NP}_{jk} (b,Q,Q_0,x_1,x_2) = 
  \exp \left[-\ln \left( Q^2\over Q^2_0 \right) h_1(b)
    -h_{j/h_1}(x_1,b)-h_{k/h_2}(x_2,b)\right].
  \label{FNPh}
  \end{eqnarray}
The functions $h_1$, $h_{j/h_1}$ and $h_{k/h_2}$ cannot be calculated using 
perturbation theory and must be measured experimentally.
Furthermore, the CSS piece $\widetilde{W}$ is evaluated at $b_*$, with
  \begin{eqnarray}
  b_* = {b \over \sqrt{1+(b/b_{max})^2} }
  \label{bStar}
  \end{eqnarray}
such that $b_*$ never exceeds $b_{max}$ \cite{berger_meng}.

To obtain the final product of our calculation,
the kinematics of the $t$ and $\bar t$, we transform the four--momentum
of $t$ ($\equiv p^\mu$) and $\bar t$ 
($\equiv {\bar p}^\mu$) from the Collins--Soper frame to the laboratory frame.
The resulting expressions are:\footnote{Our convention is $q^\mu =
(q^0,q^1,q^2,q^3)$.}
  \begin{eqnarray} & &
  p^\mu = {Q \over 2}( {q^\mu \over Q} + \sin\theta\cos\phi
X^\mu + \sin\theta\sin\phi Y^\mu + \cos\theta Z^\mu), \nonumber \\
 & & {\bar p}^\mu = q^\mu - p^\mu, \nonumber \\
 & &  q^\mu = (M_T\cosh y,Q_T \cos\phi, Q_T \sin\phi, M_T\sinh y),
\nonumber \\
 & &  X^\mu = -{Q \over Q_T M_T}(Q_{+}n^\mu + Q_{-}{\bar n}^\mu - 
{M_T^2 \over Q^2}q^\mu), \nonumber \\
 & & Y^\mu = \epsilon^{\mu\nu\alpha\beta}{q_\nu \over Q}Z_\alpha X_\beta,
\nonumber \\
 & & Z^\mu = {1 \over M_T}(Q_{+}n^\mu - Q_{-}{\bar n}^\mu),\nonumber
  \end{eqnarray}
with $Q_\pm = {1 \over \sqrt{2}}(q^0\pm q^3), Q=\sqrt{q^2}, 
M_T = \sqrt{Q^2+Q_T^2}, y={1 \over 2}\ln({Q_{+} \over Q_{-}}),
n^\nu = {1\over\sqrt{2}}(1,0,0,1),$ and ${\bar
n}^\nu={1\over\sqrt{2}}(1,0,0,-1)$.

\section{The Numerical Results of Resummation}
\indent

In this section, we present numerical results for the 
 $q \bar q \rightarrow t \bar t$ and $gg \rightarrow t \bar t$
subprocesses 
after applying the resummation formalism outlined in the previous
section. For these results, we have assumed $m_t$ = 175\,GeV 
for $t \bar t$ production at  
the Tevatron (a ${\rm p}\overline{\rm p}$ collider)
with $\sqrt{S}=1.8$\,TeV. 

As explained in the previous section, 
the CSS piece depends on the renormalization 
constants $C_1, C_2=C_1/b_0$ and $C_3=b_0$.
The choice of $C_2$ indicates that the hard scale of the
process is
$Q=C_2 M_{t \bar t}$, where $M_{t \bar t}$ is the
invariant mass of the $t \bar t$ pair.
We use CTEQ3M NLO parton distribution functions (PDF's) \cite{cteq3m}, the
NLO expression for $\alpha_s$, and
the non-perturbative function~\cite{glenn}
  \begin{eqnarray}
  F^{NP} (b,Q,Q_0,x_1,x_2) = {\rm exp} 
  \left\{- g_1 b^2 - g_2 b^2 \ln\left( {Q \over 2 Q_0} \right) - 
  g_1 g_3 b \ln{(100 x_1 x_2)} \right\},
  \label{FNPg}
  \end{eqnarray}
where $g_1 = 0.11\,{\rm GeV}^2$, 
$g_2 = 0.58\,{\rm GeV}^2$, $g_3 = -1.5\,{\rm GeV}^{-1}$ 
and $Q_0 = 1.6\,{\rm GeV}$.\footnote{
These values were fit for CTEQ2M PDF and $C_2$=1, 
and in principle should be refit for CTEQ3M PDF and different values
of $C_2$.  Also, for the $gg$ process, $g_2$ should be
replaced by $g_2 {A^{(1)}_{gg} \over A^{(1)}_{qq}} = g_2{9\over 4}$ via
the renormalization group argument for the
$\ln ({Q\over 2Q_0})$ dependence of the non--perturbative
function.  Since the $gg$ channel is
numerically less important at the Tevatron, we still use Eq.~9 in 
this study.}
Finally, the
CSS piece is fixed by specifying the order in $\alpha_s$ of the
$A,B$ and $C_{jk}$ functions.  We adopt the notation $(M,N)$ to represent
the order in $\alpha_s$ of $A^{(M)}, B^{(M)}$ and $C^{(N)}_{jk}$.
The choice $(1,0)$, for
example, means that $A$ and $B$ are calculated to order $\alpha_s$,
while $C_{jk}(z)$ is either 0 or $\delta(1-z)$ depending on $j$ and $k$.  
Also, $(1,0)$ means the $\alpha_s^3 {\cal H}_{jk}$ term in
Eq.~\ref{WTwi} is not included.

If the CSS piece is expanded to order $\alpha_s^3$, then
it contains all the $\ln({Q^2 \over Q_T^2})$ 
terms predicted by the NLO calculation.
As discussed in the previous section, the regular $Y$--piece is 
calculated by subtracting all the singular contributions, 
which grow as ${1 \over Q_T^{2}}\times (1 \, {\rm or} \, 
{\ln({Q^2 \over Q_T^2})})$ as $Q_T \rightarrow 0$,
out of the NLO tree level processes
$ q \bar q \rightarrow t \bar t g $,
$ g q/{\bar q}\rightarrow t \bar t q/{\bar q}$,
and $g g \rightarrow t \bar t g$.
In Fig.~1, we show the relative sizes of the CSS piece,
the singular--piece expanded to order $\alpha_s^3$,
and the NLO tree level result (which is also order $\alpha_s^3$) as a 
function of $Q_T$.  The regular $Y$--piece, which is defined as the 
difference
between the NLO tree level result and the singular--piece, is small for
$Q_T$ up to 50 GeV.  Its relative contribution
starts at 0 and reaches about 30\% at $Q_T$=50 GeV.  The two
curves with singular behavior as $Q_T \rightarrow 0$ have been cut off
at $Q_T$=2 GeV for the purposes of this figure only.

We conclude that the $Y$--piece is not important for small 
$Q_T$ up to about 25\,GeV, while most of the rate occurs at
much smaller $Q_T$.  In Table~\ref{tab_css}, we present the
rate for the CSS piece alone for $Q_T < 50$ GeV for each
choice of order $(M,N)$ and the dependence of this rate on the
renormalization constant $C_2$.
Results for the $gg\rightarrow t{\bar t}$ channel are only presented 
up to order (1,1).
In the same table, we show the mean and standard deviation
for each order $(M,N)$.  For the highest order, (2,1), the variation
with the hard scale set by $C_2$ 
of the $q {\bar q}\rightarrow t{\bar t}$ channel is only 2\%.  
For the $gg\rightarrow t{\bar t}$ channel, there
is only a marginal improvement in the variation from higher order, 
though the overall correction at higher order is large.  This raises
some concern regarding the higher order dependence of this result.
A similar behavior is exhibited in the $gg$ channel for the other
resummation schemes.  Fortunately, at the Tevatron, the dominant
contribution to the $t \bar t$ production rate ($\simeq 90\%$) 
comes from the stable $q {\bar q}\rightarrow t{\bar t}$ channel.
In Fig.~2, we show the relative contributions of the $q\bar q$ and
$gg$ channels to the CSS piece in this formalism.
The total rate for $t \bar t$ production is obtained by adding the
$Y$--piece to the CSS piece.  These results are compiled in 
Table~\ref{tab_sigtot} in a similar fashion as in Table~\ref{tab_css}.
We have also included a column $\alpha_s^{(N+2)} \sigma_{pert}$ which
shows the LO (for $N=0$) and NLO (for $N=1$) perturbative cross sections
for comparison.
The choice of order (1,1) for both the
$q {\bar q}\rightarrow t{\bar t}$ and $gg\rightarrow t{\bar t}$ contributions 
represents the same order
as the NLO calculation.  With the choice of hard scale $Q=M_{t \bar t}/2$ 
($C_2=1/2$),
the integrated total production rate for top quark pairs
is found to be 5.64\,pb, which agrees within 10\% with the NLO result
5.06\,pb
evaluated at the similar hard scale $Q=m_t$.
This indicates that the CSS resummation formalism presented in
the previous section
contains the dominant contribution of the NLO result to the 
production rate of $t \bar t$ pairs.  Our average resummed result
$5.58\pm0.09$~pb agrees with Ref.~\cite{berger}, which was obtained using
principal value resummation.
In the literature, the cross section for $t \bar t$ production
is usually given by
taking the hard scale to be a multiple of $m_t$ rather than $M_{t \bar t}$.
For the LO and NLO calculations and other resummation formulations
which implicitly integrate out the $Q_T$ dependence, the only
hard scale left in the problem is the mass $m_t$. 
The LO and NLO perturbative results in the final column of 
Table~\ref{tab_sigtot} are evaluated at the scale
$Q=C_2\times (2m_t)$.
The scale $Q=m_t$, then, corresponds approximately to
choosing the renormalization constant $C_2 = 1/2$ in the CSS
formalism.  

\begin{table}
\renewcommand{\arraystretch}{1.33}
\begin{center}
\begin{tabular}[bht]{|c||c|c|c|c|}\hline
Process & (M,N) & $C_2$ & $\sigma_{\rm CSS}$ (pb)& ${\bar \sigma_{\rm
CSS}}\pm\delta{\bar \sigma_{\rm CSS}}$ (pb) \\
\hline
$q\overline{q}\rightarrow t\overline{t}+X$ 
                &  (2,1)  &  1   & 4.54 & \\ \cline{3-4}
                &         &  1/2 & 4.55 &4.50$\pm$.07 \\ \cline{3-4}
                &         &  1/4 & 4.42 & \\ \cline{3-4} \cline{2-5}
                &  (1,1)  &  1   & 4.65 & \\ \cline{3-4}
                &         &  1/2 & 4.70 &4.65$\pm$.05 \\ \cline{3-4}
                &         &  1/4 & 4.60 & \\ \cline{3-4} \cline{2-5}
                &  (1,0)  &  1   & 3.64 & \\ \cline{3-4}
                &         &  1/2 & 3.93 &3.95$\pm$.32 \\ \cline{3-4}
                &         &  1/4 & 4.28 & \\ \cline{3-4} \hline
$gg\rightarrow t\overline{t}+X$ 
                &  (1,1)  &  1   & 0.81 & \\ \cline{3-4}
                &         &  1/2 & 0.78 & .77$\pm$.05 \\ \cline{3-4}
                &         &  1/4 & 0.71 & \\ \cline{3-4} \cline{2-5}
                &  (1,0)  &  1   & 0.33 & \\ \cline{3-4}
                &         &  1/2 & 0.36 & .36$\pm$.03 \\ \cline{3-4}
                &         &  1/4 & 0.39 & \\ \cline{3-4} \hline
\end{tabular}
\end{center}
\caption{CSS Contribution to the Total Cross Section for Top--Antitop 
Production at the Tevatron}
\label{tab_css}
\end{table}

\begin{table}
\renewcommand{\arraystretch}{1.33}
\begin{center}
\begin{tabular}[bht]{|c||c|c|c||c|}\hline
(M,N) for $q {\bar q}+gg$ & $C_2$ & $\sigma$ (pb)& 
${\bar \sigma}\pm\delta{\bar \sigma}$ (pb) & $\alpha_s^{(N+2)}$ 
$\sigma_{\rm pert}$ (pb) \\
\hline
(2,1)+(1,1)  &  1   & 5.51 &    & \\ \cline{2-3}
             &  1/2 & 5.49 &5.43$\pm$.12  & \\ \cline{2-3}
             &  1/4 & 5.29 &              & \\ \cline{2-3} \hline
(1,1)+(1,1)  &  1   & 5.62 &  & 4.71 \\
\cline{2-3}\cline{5-5}
             &  1/2 & 5.64 & 5.58$\pm$.09 & 5.06 \\
\cline{2-3}\cline{5-5}
             &  1/4 & 5.47 &              & 4.85 \\
\cline{2-3}\cline{5-5} \hline
(1,0)+(1,0)  &  1   & 4.13 & & 3.00 \\
\cline{2-3}\cline{5-5}
             &  1/2 & 4.45 & 4.47$\pm$.35 & 4.03  \\
\cline{2-3}\cline{5-5}
             &  1/4 & 4.83 &              & 5.57  \\
\cline{2-3}\cline{5-5} \hline
\end{tabular}
\end{center}
\caption{Total Cross Section for Top--Antitop Production 
at the Tevatron in the CSS Formalism}
\label{tab_sigtot}
\end{table}

Our resummation calculation only includes the finite
contributions from the virtual diagrams which are the same as
those in the Drell--Yan process \cite{sterman} 
for the $q {\bar q}\rightarrow t{\bar t}$ channel and those 
in Higgs production \cite{cphiggs} for the $gg\rightarrow t{\bar t}$
channel,
plus terms containing the running of $\alpha_s^2$ in the hard part of
the cross section multiplying ${\cal H}_{jk}$.
Since there is no
final state QCD radiation in the Drell--Yan or Higgs production
processes, this is clearly
an approximation.  If the exact virtual corrections were included,
the Wilson coefficient functions $C_{jk}$ would have to be modified in
a manner consistent with the CSS formulation, i.e.
assuming that the initial and final
state gluon radiation for the $t \bar t$ process factorizes 
in a similar fashion as the initial
state radiation alone in the Drell--Yan process.
While such a factorization is reasonable, there is no formal proof.
Therefore, we approximate the $q{\bar q} \rightarrow t {\bar t}$ 
($gg\rightarrow t{\bar t}$) finite virtual corrections with
those from the Drell--Yan (Higgs production) process.
Furthermore, the factor $B^{(2)}$ for $q{\bar q} \rightarrow t {\bar t}$ 
inside the Sudakov factor
is likely to be different from that for the Drell--Yan process,
so we did not include it in our calculations.
As shown in
Tables~\ref{tab_css} and~\ref{tab_sigtot},
as $C_2$ varies between $1/4$ and 1, the total rate varies
by only a few percent, which is about the same magnitude as
the uncertainty in the NLO calculations.
Clearly, all the other finite contributions from
the virtual diagrams other than those similar
to the Drell--Yan virtual contributions for 
$q {\bar q}\rightarrow t {\bar t}$ are small~\cite{smith_approx}.
Since $A^{(2)}$ for this process is the same as for
the Drell--Yan process, we have included it in our order (2,1) calculation 
for $q{\bar q} \rightarrow t {\bar t}$ 
to improve our predictions, which yields the results shown in the
first row of Table~2.

Because of the agreement of our predictions for the total rate with the
NLO calculation,
we apply the results of our calculations to study
the kinematic distributions of the $t \bar t$ pair and the
individual $t$ or $\bar t$ produced in hadron collisions.
In the next section, we present these kinematic distributions
and compare them to those predicted by showering event generators.
We wish to stress that this is the only type of comparison that is
sensible, because the LO calculation predicts a $\delta(\vec{Q}_{T})$
dependence for the $Q_T$ distribution, the NLO calculation cannot
accurately describe the $Q_T \ll m_t$ region of phase space, 
and other resummation
formalisms integrate out the $Q_T$ dependence and, thus, cannot
predict the kinematic distributions of $t$ and $\bar t$.

In the following sections, we present results only for order $(2,1)$ and
the canonical choice of renormalization constants $C_2$=1.
There are at least two reasons for doing this.  First, the coefficients
of the non--perturbative function used in this study were fit to data
assuming $C_2=1$.  While the non--perturbative function can affect the
shape of distributions, it does not affect the total rate.  Therefore, the
stability of our results to variation of $C_2$ is still valid, but we
cannot trust other choices of $C_2$ to give the correct shape.
Second, since we do not integrate out the kinematics, we cannot argue
that the only scale left in the problem is $m_t$.  Since an s--channel
process dominates, $Q=M_{t\bar t}$ is motivated by the dynamics.

We have also checked the effect of neglecting final state radiation, 
and have found
that it reduces the total contribution of the
CSS piece to $t \bar t$ production by about 
10\% in our approximation and slightly changes the shape near the
peak.  Without the final state radiation, however, the $Y$--piece is
not finite as $Q_T\rightarrow 0$; therefore we include it in our
results.

\section{Comparison with the Showering Monte Carlo Technique}
\indent

In this section, we compare our resummed results for the kinematics
of the $t \bar t$ pair and the individual $t$ or $\bar t$
with those predicted
by the showering event generator {\tt PYTHIA}.  To make this comparison
more understandable, we first explain the approximate implementation of
resummation in such generators.  The starting point for the
showering Monte Carlo technique is the observation that the leading
logarithmic singularities in NLO calculations are contained in the
Altarelli--Parisi splitting kernels.  In this leading log approximation,
successive parton emissions occur independently, modulo some angular
ordering effects.  In a Monte Carlo simulation, which has explicit
finite cutoffs for the energy of the emitted radiation, it is possible
to treat these emissions as a Markov chain stretching 
from the hard scattering {\it backwards} to the initial state partons
\cite{sjo85}.  
Essentially, one chooses the
kinematics for a process at the hard scale $Q_{max}$
based on the LO cross section, then
calculates the
probability for no parton emission in evolving from a high scale
$t_{max}\approx \ln(Q_{max}^2/\Lambda_{QCD}^2)$ to a lower scale $t$.
This probability is given in the leading log approximation by
the Sudakov form factor $e^{-S(x,t,t_{max})}$, where
\begin{eqnarray*}
  S(x,t,t_{max})  = \int_{t}^{t_{max}} d\acute{t} \sum_{a}^{}
  \int_{}^{} {d\acute{x}  \over \acute{x}} { \alpha_s(\acute{t}) \over 2\pi }
   {f_a(\acute{x},\acute{t}) \over f_b(x,\acute{t})} 
    P_{a \rightarrow bc} \left ({x \over \acute{x}}\right ), \\
           ~~~  = \int_{t}^{t_{max}} d\acute{t} \sum_{a}^{}
  \int_{}^{} {dz} { \alpha_s(\acute{t}) \over 2\pi } 
  {\acute{x}f_a(\acute{x},\acute{t}) \over x f_b(x,\acute{t})} 
  P_{a \rightarrow bc}(z).
\end{eqnarray*}
Here, $f_a(x,t)$ and $f_b(x,t)$ are parton distribution functions 
for partons $a$ and $b$
and $P_{a \rightarrow bc}(z)$ is the Altarelli--Parisi splitting function
for the branching $a\rightarrow bc$ with momentum fraction $z$.
In improved treatments of the Sudakov form factor, such as in {\tt
PYTHIA}, $\alpha_s$ is evaluated not at the scale $Q^2$, but at
$Q^2(1-z)$ \cite{bas83}.
If no radiation occurs down to some cutoff $t_{min}$,~$\approx$~2~GeV in 
{\tt PYTHIA}, then the parton is placed on the mass shell.
On the other hand, if it is determined that radiation does occur before
the cutoff, then a new parton is emitted.
The type of branching $a\rightarrow bc$ which led to the parton emission
is determined by the relative weights of the $x^{'}$ integrals over
$P_{a\rightarrow bc}$. 
The initiator of the
splitting is given the virtuality $t$, and the process continues until
a parton reaches the scale $t_{min}$, and it is placed on the mass shell.
With each splitting, the remaining kinematics are sampled so that energy and
momentum are conserved.  
The result is a total cross section given by the LO calculation, but
with the kinematic distributions of a resummed calculation.  
In this sense, it corresponds to a choice of $(M,N)=(1,0)$ in the
CSS formalism.\footnote{The reason there is not exact agreement between
the LO and (1,0) rates in Table~2 is because of the explicit cutoff
$b_{max}$ in the Fourier transform.}
Final state radiation is implemented in a similar fashion, but it
occurs {\it forward} from the hard scattering process to the final state
and is not weighted by the parton distribution functions.

First, we present a comparison of the resummed and showering Monte Carlo
kinematics for the $t \bar t$
pair.  Several observables which in principle cannot be extracted from a
NLO calculation are the distributions of the transverse momentum of
the $t \bar t$ pair $Q_T$, the opening angle between $t$ and $\bar t$ in
the azimuthal plane $\Delta\phi_{t \bar t}$,
and the variable $z\equiv-{ {\vec p_{T}}(t) \cdot {\vec p_{T}}({\bar t}) \over 
|max({p_{T}^2}(t),{p_{T}^2}({\bar t}))|}$.  
The resummed results are separately shown as solid lines.
Displayed on the same plots are the shapes predicted by { \tt PYTHIA} for
two choices of hard scale, $Q^2= \hat{s}$ and $Q^2= \sqrt{m_t^2+p_T^2}$, where
$p_T$ is the transverse momentum of the top quark at LO.
As illustrated in Table~2, the LO rate is highly sensitive to the
choice of hard scale.  The resummed estimate of the total rate is
more reliable.
Therefore, to compare the shapes of distributions, we
have renormalized the {\tt PYTHIA} results to have the same total rate
as the resummed calculation.

Fig.~3 shows the distribution of the resummed variable $Q_T$.  
Note that the resummed
distribution is significantly harder than the showering Monte Carlo
result, implying that the resummed calculation predicts more overall
hard radiation.  
This is further demonstrated by the $\Delta\phi_{t\bar t}$ distribution,
the difference in azimuthal angle between $t$ and $\bar t$, in
Fig.~4, which is depleted near $\Delta\phi=\pi$ in comparison to the
showering Monte Carlo.  
Because of initial or final state radiation, the $t$ and $\bar t$
are not expected to be exactly back--to--back.
The $z$ distribution, shown in Fig.~5, is also shifted
away from a back--to--back configuration $(z=1)$ in the resummed result.
Finally, the distributions of the rapidity of the 
$t \bar t$ pair $y_{t\bar t}$ and the invariant mass $M_{t\bar t}$
are shown in Figs.~6 and 7 respectively.
The rapidity distribution
is more central for the resummed result 
and the invariant mass favors higher masses and is broader.
These differences arise
because {\tt PYTHIA} only contains LO matrix elements, while the
resummed 
result contains the dominant piece of the NLO correction.

Second, we present a comparison of the kinematics of the individual 
$t$ or $\bar t$.
Fig.~8 shows the distribution of the transverse momentum of the
individual $t$ $p_T^t$, while Fig. 9 shows
the difference in their rapidity $\Delta y^t$, where
$\Delta y^t=y^t-y^{\bar t}$.  
From the LO calculation, we know the scale of $p_T^t$ is set by 
$m_t$ ($p_T^t\simeq m_t/3 \simeq$
60 GeV), while the typical transverse momentum $Q_T$ is much smaller.
Therefore, there is not much difference in these distributions.
Likewise, $\Delta y^t$ is more sensitive
to the PDF (which determines the boost of the $t\bar t$ pair)
than the transverse momentum $Q_T$, so we do not
expect to observe a large difference.  
In conclusion, the showering generators
(as typified by {\tt PYTHIA} in our study) reproduce the CSS
distributions for the individual $t$ and $\bar t$ and the $t\bar t$ pair
kinematics
in our plots to a 10\% level bin--by--bin,
although the overall shapes are generally different and the complete
resummed results indicate more overall hard radiation.
Based on these results, we seek to improve the showering Monte Carlo
technique by using our knowledge of the resummed $Q_T$ distribution.
This is discussed in the next section.

\section{Jet Activity in $t\bar t$ Events at Hadron Colliders}
\indent

Despite its limitations in predicting the correct rate, the showing
Monte Carlo technique has a mechanism to approximate the complete
resummed result for the $Q_T$ distribution.  Furthermore, the
showering event generator gives a phenomenologically accurate
description of all the details of the event.
The resummation calculation only predicts the vector sum of all
soft gluon radiation, but has no power to predict how the radiation
is distributed amongst individual gluons or quarks.  Such details are crucial
for estimating the amount of jet activity, which will affect the
determination of the top quark mass by reconstructing jets from
top quark decays as done by CDF and D0 at the Tevatron \cite{top}.  
These estimates are
used to engineer cuts to enhance the signal and to optimize the
choice of jet definition.  In this section, we present a simple
synthesis of the showering Monte Carlo technique and the full
resummation calculation to realize a better estimate of the hadronic 
activity.  This synthesis is accomplished by reweighting events generated
by {\tt PYTHIA} to agree with the resummed $Q_T$ distribution and rate.

To demonstrate the improved predictive ability of the resummed approach,
we present the distributions of the
jet multiplicity and scalar sum of the transverse
energy in $t \bar t$ events.  We have isolated the contributions from
initial and final state radiation only, so that none of the the $t$ or
$\bar t$ decay products (or additional QCD radiation contributions from
them) are included in these plots.  We define jets using a simple
calorimeter simulation (with cell segmentation 
$\Delta\eta\times\Delta\phi = .1\times.1$) and a
clustering algorithm based on a cone size $R=.4$ and a minimum jet
transverse energy $E_T^{min}$=5 GeV.  The energy $E$ deposited in each
cell of the calorimeter is smeared with a Gaussian resolution
${\sigma_E \over E} = {.70 \over \sqrt{E}}$.  The jet multiplicity
distribution is shown in Fig.~10.  Note the shift in the peak value
of the distribution from 0 jets to 1 jet.  The resummation based
Monte Carlo clearly predicts more hard radiation in $t \bar t$ events at
hadron colliders.  Similar information is conveyed in Fig.~11, which
shows the scalar sum of the transverse energy of the jets defined
as above.
While the details of jet observables can only be studied using the showering
Monte Carlo technique, our knowledge of the resummed $Q_T$ distribution
allows improvement.  
It is straightforward to extend these results to include the decay products
of the $t$ and $\bar t$ and even the hard gluons from the QCD radiative
decay of the top quark \cite{qcdrad}.  We shall leave this for further
study.

As a final point of comparison, we have attempted to quantify the effect
of hard gluon radiation on the extraction of $m_t$ from data.
The issue at hand is how often a hard gluon from radiation is
misidentified as a top quark decay product.  Using a sample of
events where $t(\to b W^{+}(\to e^{+}\nu_e)) {\bar t}(\to{\bar b}
W^{-}(\to jj))$, we cluster particles into jets as described above
and analyze events with 4 or more hard jets ($E_T^{j}>$15 GeV,
$|\eta^{(j)}|<2$).  
We separate the jets than can be identified with
a $\bar t$ decay product (for simplicity, we assume the $b$ is
correctly tagged) and find the one with the lowest $E_T$.
The fraction of events where another jet (i.e. one not from $\bar t$ decay) 
has a higher $E_T$ than this is an estimate of the importance of
the hard radiation.  While this is only a crude estimate of the
real effect, we find that the fraction ($= .56$) does not differ
between the standard and improved {\tt PYTHIA}.  This implies
that the hard gluon error on the top mass measurement is well
estimated by showering generators, though this requires further 
study.  We have not attempted to quantify the effect of
soft gluon radiation, where the radiation is not resolved as
an individual jet but can overlap with the $t\bar t$ decay products,
though the complete resummed result indicates that this too should
be enhanced.

\section{Conclusions}
\indent

To further study the interactions of the top quark and
to better measure its mass,
we must understand the kinematic distributions of 
the transverse momentum, rapidity and azimuthal angle 
of the top quarks. 
The kinematics of the top quarks produced at hadron colliders can 
be accurately predicted only after resumming the multiple soft gluon 
emissions in either the initial state or the final state.
In this work, we have adopted the CSS resummation formalism to
obtain the kinematics of the top quarks.                          
The approximation we made in this study should be adequate because
of the large top quark mass. The important consequence of the large
top quark mass is that
the logarithmic terms from the initial state are more important than
those
from the final state. In the former case, there is 
double--logarithmic behavior due to both soft and collinear
gluon radiation, while,
in the latter case, only single-logarithmic behavior due to soft but not
collinear gluon radiation is 
possible up to the order
$\alpha_s^3$.
We compared our resummation calculation with those predicted by the 
full event generator {\tt PYTHIA} and found that the latter
does not give the same prediction as ours. 
In view of the fact that the full event generator has been widely used by our 
experimentalist colleagues for analyzing their data, it
is important to point out the difference between the results from 
this approach and those from the analytical calculation.
In determining the mass of the top quark, for example, one needs to
account for the jet activity in at least two cases: (1) when a hard
jet from the showering is misidentified as one of the top decay jets,
and (2) when the soft radiation is included in the energy determination
of true top decay products.
The hybrid approach presented in Section~5, which relates the transverse 
momentum of
the top--antitop quark pair from an analytic calculation to the 
balancing gluon radiation from a showering Monte Carlo, 
contains a more realistic description of event structure,
which could be used for choosing kinematic cuts and tuning
the jet energy correction algorithm.    

\section*{ Acknowledgments }
\indent

We thank Ed Berger for starting us on
this project and providing helpful comments.
C.-P.~Y. also thanks the CTEQ collaboration and C.~Balazs for
discussions.
This work was supported in part by NSF grant PHY-9309902
and by DOE grant
DE--FG03--92--ER40701.

\newpage
\begin{figure}
\centering
\epsfxsize=3.in
\hspace*{0in}
\epsffile{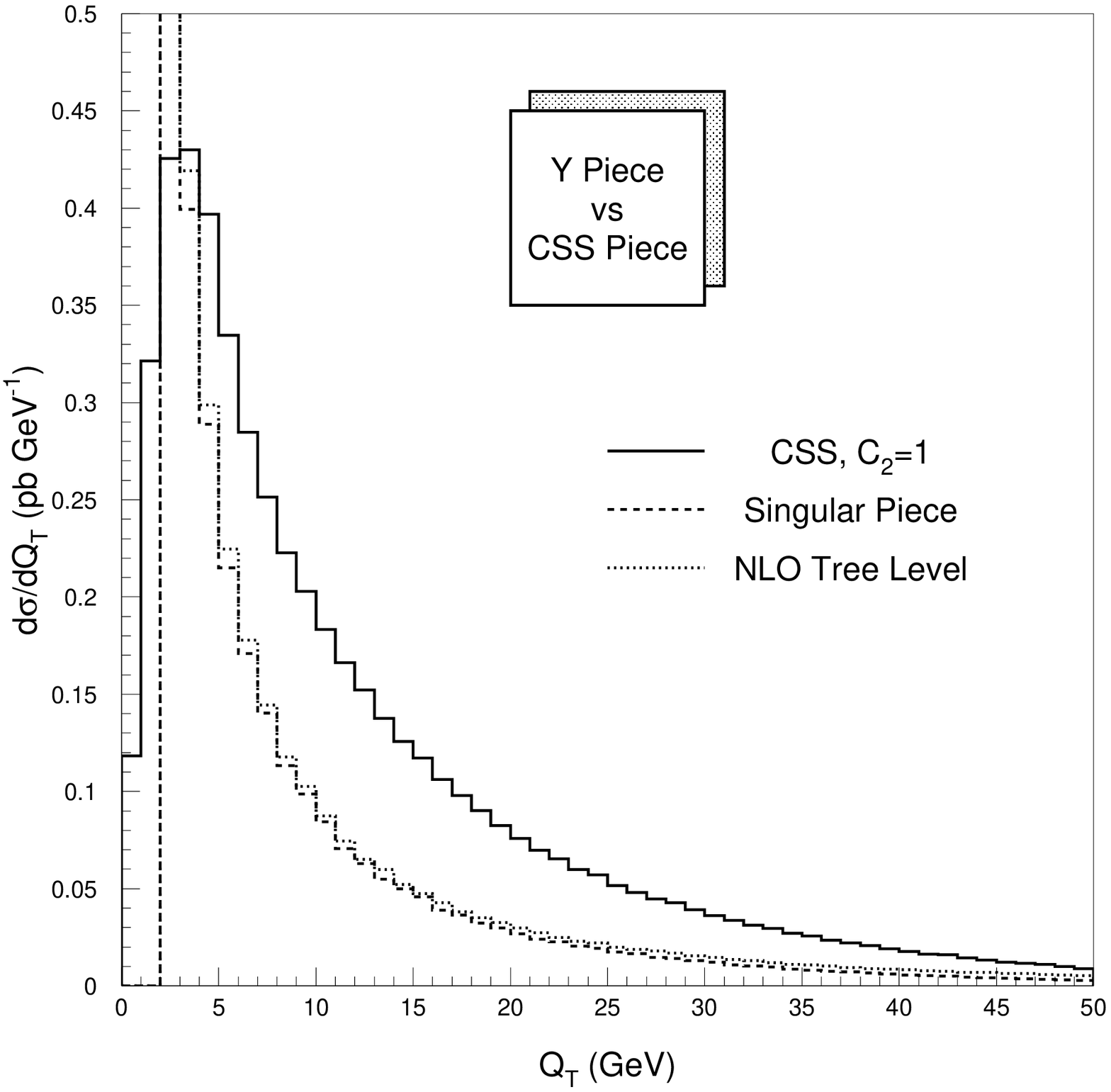}
\vspace*{0in}
\caption{The relative importance of the $Y$-piece with respect to the
CSS piece as a function of
$Q_T$.}
\label{fig_pma}
\end{figure}

\newpage
\begin{figure}
\centering
\epsfxsize=3.in
\hspace*{0in}
\epsffile{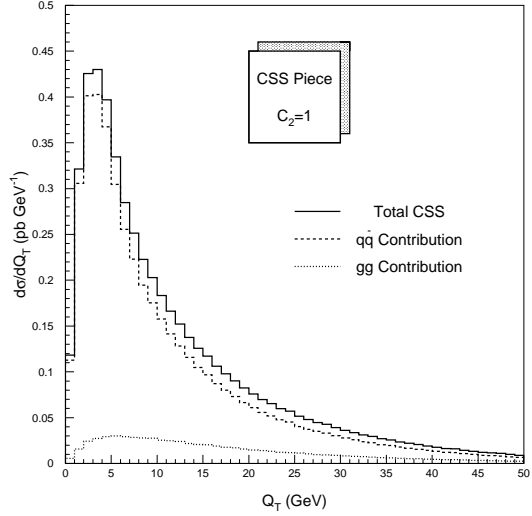}
\vspace*{0in}
\caption{The relative contributions of $q\bar q\to t\bar t$ and
$gg\to t\bar t$ to the
CSS piece as a function of
$Q_T$.}
\label{fig_qqgg}
\end{figure}

\begin{figure}
\centering
\epsfxsize=3in
\hspace*{0in}
\epsffile{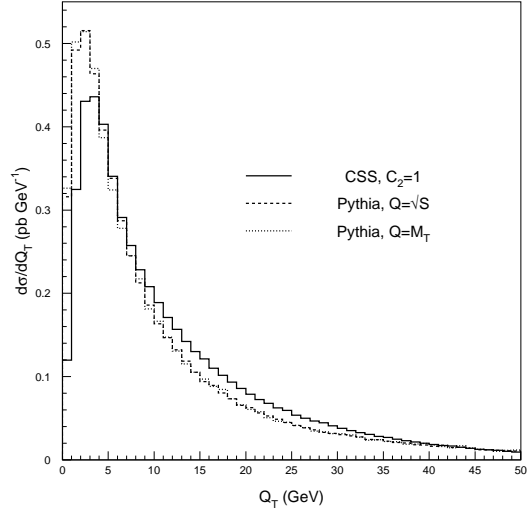}
\vspace*{0in}
\caption{The transverse momentum of the ${t \bar t}$ pair $Q_T$ for
$C_2$=1 and for different choices of hard scale for the showering
Monte Carlo {\tt PYTHIA}.  The {\tt PYTHIA} rate has been renormalized
to the CSS rate.}
\label{fig_qt}
\end{figure}

\begin{figure}
\centering
\epsfxsize=3.0in
\hspace*{0in}
\epsffile{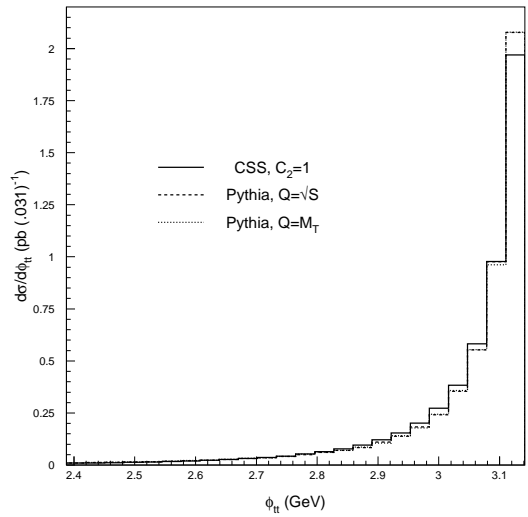}
\vspace*{0in}
\caption{The difference in azimuthal angle between the $t$ and
$\bar t$ $\Delta\phi_{t\bar t}$.}
\label{fig_phitt}
\end{figure}

\begin{figure}
\centering
\epsfxsize=3.0in
\hspace*{0in}
\epsffile{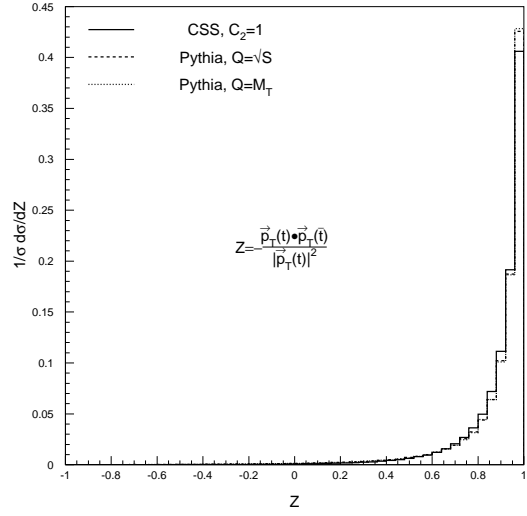}
\vspace*{0in}
\caption{$z$ distribution of the $t$ and $\bar t$.}
\label{fig_ztt}
\end{figure}

\begin{figure}
\centering
\epsfxsize=3.0in
\hspace*{0in}
\epsffile{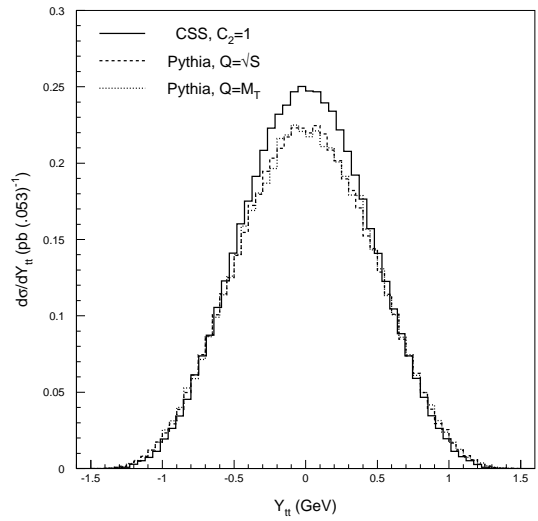}
\vspace*{0in}
\caption{Rapidity of the $t\bar t$ pair $y_{t\bar t}$.}
\label{fig_ytt}
\end{figure}

\begin{figure}
\centering
\epsfxsize=3.0in
\hspace*{0in}
\epsffile{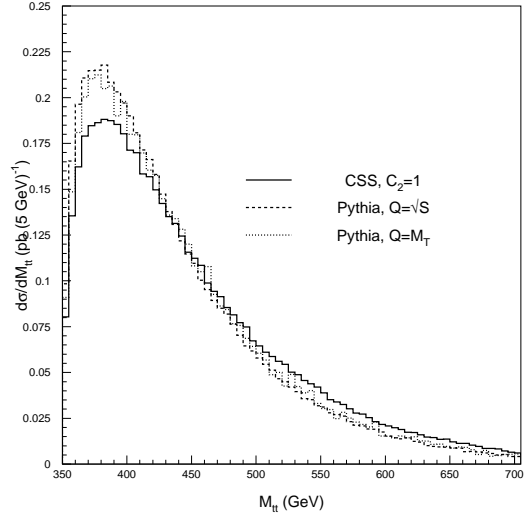}
\vspace*{0in}
\caption{Invariant mass of the $t\bar t$ pair $M_{t\bar t}$.}
\label{fig_mtt}
\end{figure}

\begin{figure}
\centering
\epsfxsize=3.0in
\hspace*{0in}
\epsffile{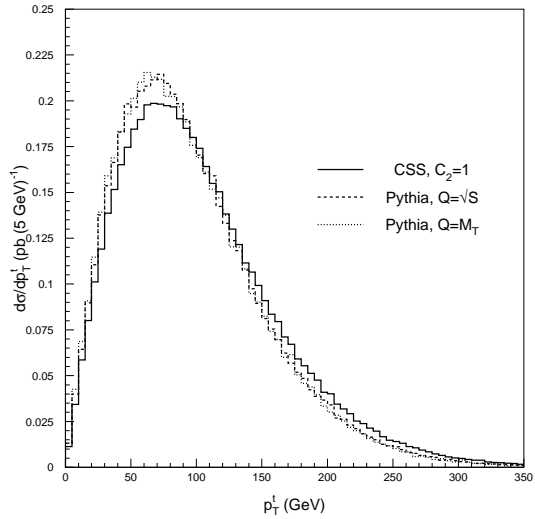}
\vspace*{0in}
\caption{Transverse momentum of the individual $t$ or $\bar t$ 
$p_{T}^{t}$.}
\label{fig_pttop}
\end{figure}

\begin{figure}
\centering
\epsfxsize=3.0in
\hspace*{0in}
\epsffile{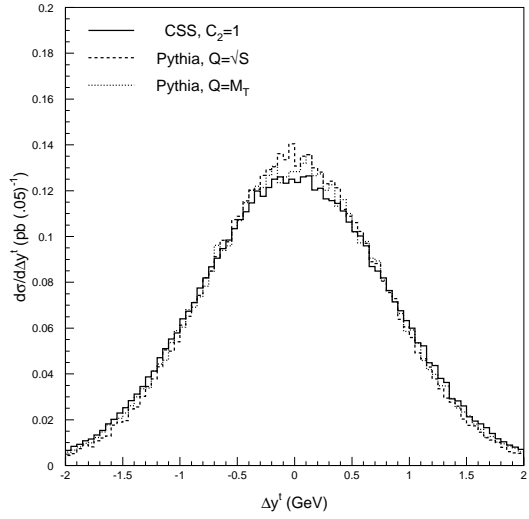}
\vspace*{0in}
\caption{Rapidity difference between the $t$ and $\bar t$ $\Delta y_t$.}
\label{fig_ytop}
\end{figure}

\begin{figure}
\centering
\epsfxsize=3.0in
\hspace*{0in}
\epsffile{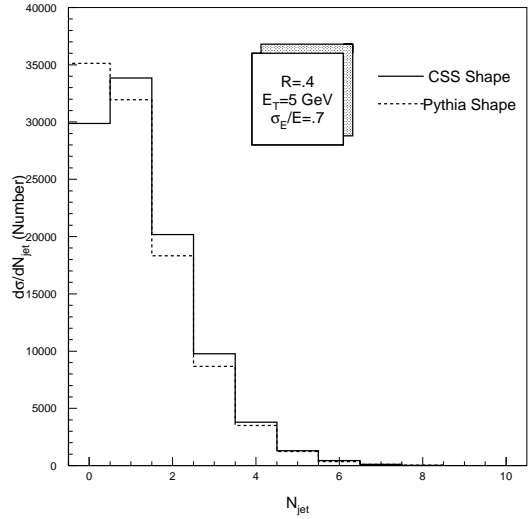}
\vspace*{0in}
\caption{Jet Multiplicity from initial and final state radiation as
predicted by {\tt PYTHIA} (dashed line) and the hybrid {\tt PYTHIA}--
CSS resummation (solid line).}
\label{fig_njet}
\end{figure}

\begin{figure}
\centering
\epsfxsize=3.0in
\hspace*{0in}
\epsffile{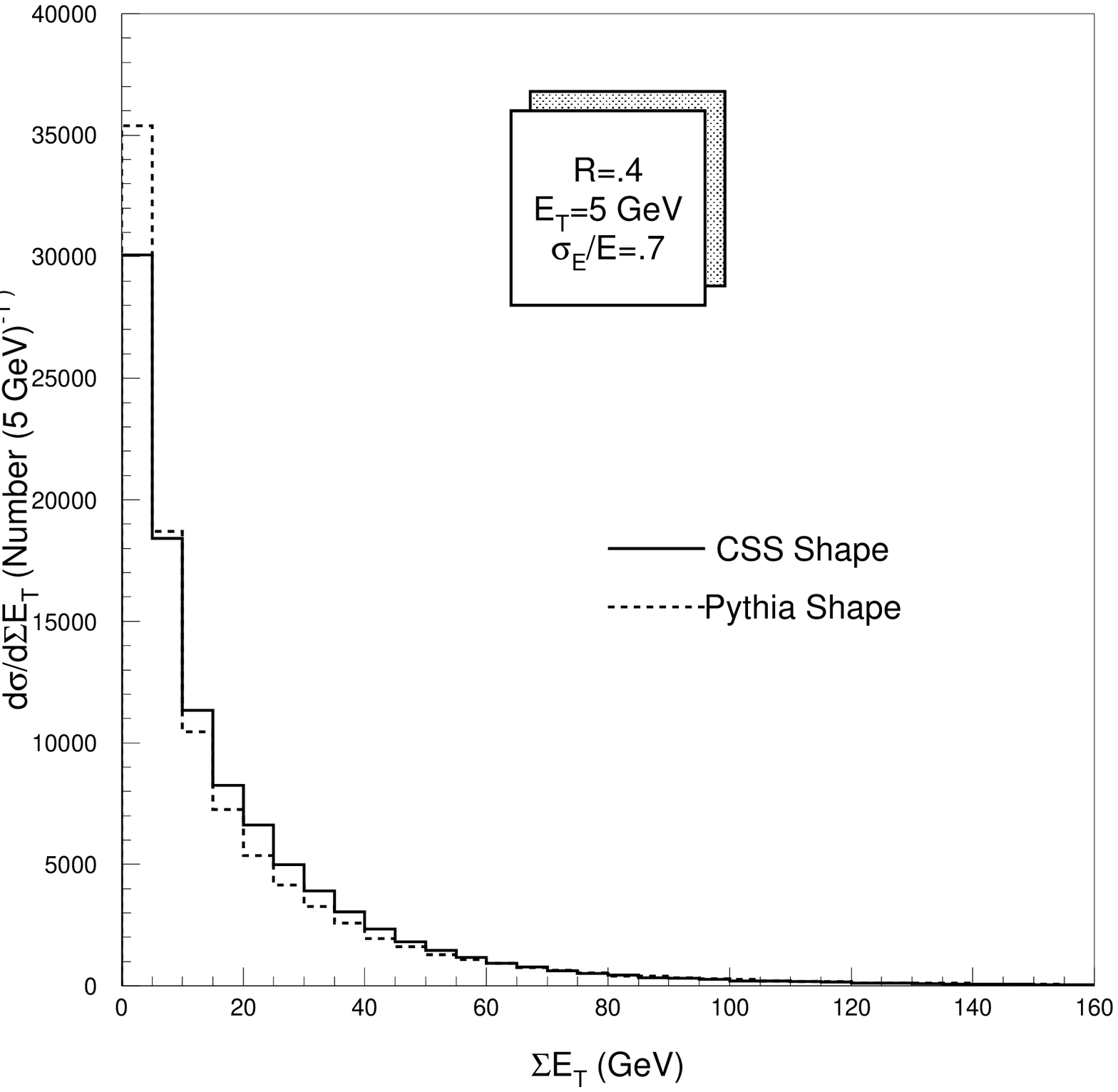}
\vspace*{0in}
\caption{Scalar sum of jet transverse energy $\sum_{}^{}E_T$ from
initial and final state radiation as
predicted by {\tt PYTHIA} (dashed line) and the hybrid {\tt PYTHIA}--
CSS resummation (solid line).}
\label{fig_sumet}
\end{figure}

\vspace{0.0cm}                                                                 
\noindent                                                                      

\begin{thebibliography}{99}
\bibitem{top}
F.~Abe {\it et al.}, Phys. Rev. Lett. {\bf 73}, 225 (1994);
\\
S.~Abachi {\it et al.}, Phys. Rev. Lett. {\bf 72}, 2138 (1994).
\bibitem{ehab}
Ehab Malkawi and C.--P. Yuan, Phys. Rev.
{\bf D50}, 4462 (1994). 
\bibitem{sekh}R.S. Chivukula, E. Gates, E.H. Simmons and J. Terning, 
Phys. Lett. {\bf B311,} 157 (1993);
\bibitem{nlotop}
P. Nason, S. Dawson and R.K. Ellis, 
Nucl.~Phys. {\bf B303}, 607 (1988); {\bf B327}, 49 (1989); \\
W. Beenakker, H. Kuijf, W.L. van Neerven and J. Smith, 
Phys.~Rev. {\bf D40}, 54 (1989); \\
R. Meng, G.A. Schuler, J. Smith and W.L. van Neerven, 
Nucl.~Phys. {\bf B339}, 325 (1990).
\bibitem{smith}
E. Laenen, J. Smith and W.L. van Neerven,
Nucl. Phys. {\bf B369}, 543 (1992).
\bibitem{berger}
E.L. Berger and H. Contopanagos, Phys.~Lett. {\bf B361}, 115 (1995).
\bibitem{mangano}
S. Catani, M.L. Mangano, P. Nason and L. Trentadue,
CERN-TH/96-21, January 1996.
\bibitem{altarelli}
G. Altarelli, R.K.~Ellis, M.~Greco, G.~Martinelli, Nucl.~Phys.~{\bf
B246}, 12 (1984).
\bibitem{berger_meng}
E.L. Berger and R. Meng, Phys. Rev. {\bf D49}, 3248 (1994).
\bibitem{collins} 
J. Collins and D. Soper, Nucl. Phys. {\bf B193} 381, (1981);
Erratum {\bf B213} (1983) 545; {\bf B197}, 446 (1982).
\bibitem{sterman} 
J. Collins, D. Soper and G. Sterman, Nucl. Phys. {\bf B250}, 199 (1985).
\bibitem{generators}
G.~Marchesini, B.R.~Webber, G.~Abbiendi, I.G.~Knowles, M.H.~Seymour,
and L.~Stanco, Comp.~Phys.~Commun.~{\bf 67}, 465 (1992);
H.~Baer, F.~Paige, S.~Protopopescu, and X.~Tata, "Simulating
Supersymmetry with ISAJET 7.00," FSU--HEP--930329, 
SSCL--Preprint--441 (1993);
T.~Sj\"{o}strand, Comp.~Phys.~Commun.~{\bf 82}, 74 (1994).
\bibitem{frixione}
S. Frixione, M.L. Mangano, P. Nason, and G. Ridolfi,
Phys. Lett. {\bf B351}, 555 (1995).
\bibitem{arnold}
P.B. Arnold and R.P. Kauffman, Nucl. Phys. {\bf B349}, 381 (1991).
\bibitem{cphiggs}
C.-P. Yuan, Phys. Lett. {\bf B283}, 395 (1992).
\bibitem{kauffman_higgs}
R.P. Kauffman, Phys. Rev. {\bf D45}, 1512 (1992).
\bibitem{CSFrame} 
J. Collins and D. Soper, Phys. Rev. {\bf D16}, 2219 (1977).
\bibitem{cteq3m}
H.L. Lai, J. Botts, J. Huston, J.G. Morfin, J.F. Owens, J.W. Qiu, 
W.K. Tung, H. Weerts,
MSU preprint MSU-HEP-41024, Oct 1994.
\bibitem{glenn} 
G.A. Ladinsky and C.-P. Yuan, Phys. Rev. {\bf D50}, 4239 (1994).
\bibitem{sjo85}
T.~Sj\"{o}strand, Phys.~Lett.~{\bf 157B}, 321 (1985).
\bibitem{bas83}
A.~Bassetto, M.~Ciafaloni and G.~Marchesini, Phys.~Rep.~{\bf 100}, 202
(1983).
\bibitem{smith_approx}
R.~Meng, G.A.~Schuler, J.~Smith, and W.L.~van~Neerven, Phys.~Lett.~{\bf
B339}, 325 (1990).
\bibitem{qcdrad}
S.~Mrenna and C.-P.~Yuan, Phys. Rev. {\bf D46}, 1007 (1992).
\end{thebibliography}
\end{document}